| | |
|---|---|
| Abstract | The available diagnostics for atmospheric micro-plasmas remain limited and relatively complex to implement; so we present a radio-frequency technique for diagnosing a key parameter here. The technique allows one to estimate the dependencies of the electron density by measuring the RF-impedance of the micro-plasma and analyzing it with an appropriate equivalent circuit. This technique is inexpensive, can be used in real time and gives reasonable results for argon and helium DC micro-plasmas in holes over a wide pressure range. The electron density increases linearly with current in the expected range consistent with normal glow discharge behavior. |

# 1. Introduction

Atmospheric pressure plasmas and in particular microplasmas are quickly increasing in importance and in numbers of applications [1]. They are used or are being considered for use in processing materials [2], in advanced lighting applications [3], for plasma thrusters [4], for various biological applications [5] including sterilization and enhanced wound healing, for ozone production and NOx abatement [1] to name just a few. These glows come in a wide variety of shapes and forms and are produced by a wide variety of means from DC power to microwave and even optical. They all seem to have in common a more extreme environment than lower pressure non-thermal plasmas and consequently increased difficulty in measuring the conditions of that environment.

Several authors have found that RF-impedance measurements of low pressure plasmas can be valuable in estimating their characteristics when performed carefully and analyzed using a reasonable model of the sheath and bulk glow [6,7]. We show in this article that the same can be applied successfully to understanding atmospheric microplasmas. Indeed, the high pressure condition helps to simplify the analysis. The technique can yield spatially integrated information that would otherwise be much more costly, difficult and time consuming to obtain.

# 2. Experimental setup

A schematic of the experimental apparatus used in this work is shown in Figure 1. Figure 1a shows the overall system used for establishing and measuring the DC micro hollow cathode discharge sources (MHCD sources) while Figure 1b details the matching network and "tuning" circuits. Figure 1a shows that the apparatus consists of: (1) a MHCD source. The source used was 0.5 mm thick Al2O3 with 8 µm thick Ni films on both sides and an approximately 180 µm diameter through-hole. The DC micro-plasma is formed in this hole and visibly expands over the cathode surface, but not over the anode surface; (2) a DC power supply for establishing the plasma. This DC power supply is connected directly to the cathode through a ballast resistor, $R_B$ (either 20 or 50 kΩ); (3) an RF power supply for applying the diagnostic voltage. We have used a signal generator (13.56 MHz) and amplifier (Agilent 33120A and ENI 601L); this RF signal is fed through (4) a matching network simply to control the mixing of RF and DC; (5) an RF "tuning circuit" is placed immediately adjacent to the MHCD source and used to establish a large RF input impedance when no plasma is present; (6) calibrated

RF I – V probes are used to measure the RF I, V and phase. The I – V probe pair consisted of an Ion Physics Corp. CM-10-MG current probe integrated into a single package with a capacitive voltage pickup. The RF signals were recorded using a Tektronix TDS 640A oscilloscope and then analyzed using LabVIEW's FFT function and the calibration data; (7) finally, two HP 34401A multi-meters are used to measure the DC voltage across the micro-plasma and the DC current flowing through it.

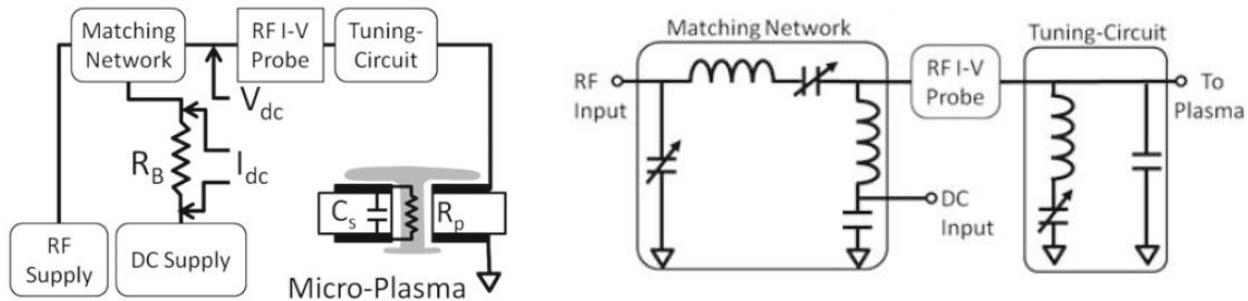

*Fig. 1. A schematic diagram of the experimental apparatus. (a) The DC and RF circuitry. (b) Detail of the RF and DC mixing as well as tuning circuit.*

Figure 1b details the RF matching network and "tuning" circuits. A matching network has been used only because it provided a convenient place to mix the DC with the RF. Because we used a stock matching network (Daihen, RMN-10) we used a 13.56 MHz RF signal as our probe frequency. The measurements do not depend upon the actual settings of the matching network in any fashion, even though the matching network can be used to control the amplitude of the RF if so desired. The matching network used in this article is a simple L-Network with variable shunt and series capacitances. The presence of the series capacitor prevents the large DC voltages needed to operate the micro-plasma from entering the RF power supply. Likewise, the shunt inductor protects the DC supply from the applied RF signal even though this signal is purposely made small. This shunt inductor is a very small DC impedance and as a consequence allows the DC power supply to strike and maintain the micro-plasma.

The "tuning" circuit is placed immediately between the RF I – V probes and the micro-plasma source. It forms a tank circuit with the micro-plasma source which can be tuned using a single variable capacitance. The objective of this circuit is to place a "tuned" inductance in parallel with the capacitance of the MHCD source ($C_s$) such that the RF input impedance measured by the RF I – V probes becomes as large as possible when there is no plasma. This tuning circuit is needed because the capacitance of the MHCD source, $C_s$, is in parallel with the plasma impedance and is much smaller than the plasma impedance. As a consequence, there would be almost no detectable difference between plasma-on and plasma-off without this tuning network properly set to resonance. The larger the input impedance without plasma, the larger the plasma impedance (smaller the charge density) one can reliably measure. This will be explained in greater detail in the analysis section. Using this tuning circuit, we were able to obtain an input impedance (without plasma) greater than 20 kΩ for all our measurements. (Typical input impedances ranged between 20 and 24 kΩ depending mostly upon the accuracy with which we were able to set the tuning capacitance.) Because small changes in source temperature can cause changes in the micro-plasma source capacitance (through thermal expansion etc.) the tuning circuit was re-adjusted for each measurement. In addition, the input impedance was measured without plasma before and after each measurement. While the input impedance of the source itself was always capacitive, it was not large enough in capacitance to facilitate such a tank circuit. (We needed too large an inductance to balance the capacitance, $C_s$, of the source by itself.) Consequently, we added another capacitance in shunt with the MHCD source at the output of the tuning circuit (as shown in Fig. 1b). This capacitance allowed us to tune the input impedance of the MHCD source without plasma to a local maximum.

It is important that both the RF power delivered to the micro-plasma and the RF voltage be much smaller than the DC power and voltage used to produce the plasma so that the RF does not change the DC plasma conditions significantly. This allows one to measure the DC plasma characteristics rather than the characteristics of mixed RF-DC plasmas. Achieving such conditions (RF power << DC power) is not difficult at all. Figure 2 shows the power delivered to Argon micro-plasmas at 4 different pressures. The DC power is approximately linearly dependent upon the DC current through the plasma since the DC I – V characteristic has the discharge voltage independent of the discharge current. (The DC discharge operating voltage is ~200 V.) We note that the DC power ranges from approximately 0.6 W to greater than 3 W in this single hole. In contrast, the RF power required to accurately measure these glows was kept below 7 mW. In fact, in the measurements presented here, the RF power was always less than 0.5% of the DC power and, the RF Voltage was always less than 5% of the DC Voltage as well. In the RF power curves shown in Figure 2b, we note that the RF measurements at 1000 Torr were made at lower power to show that the power could be reduced even more if desired.

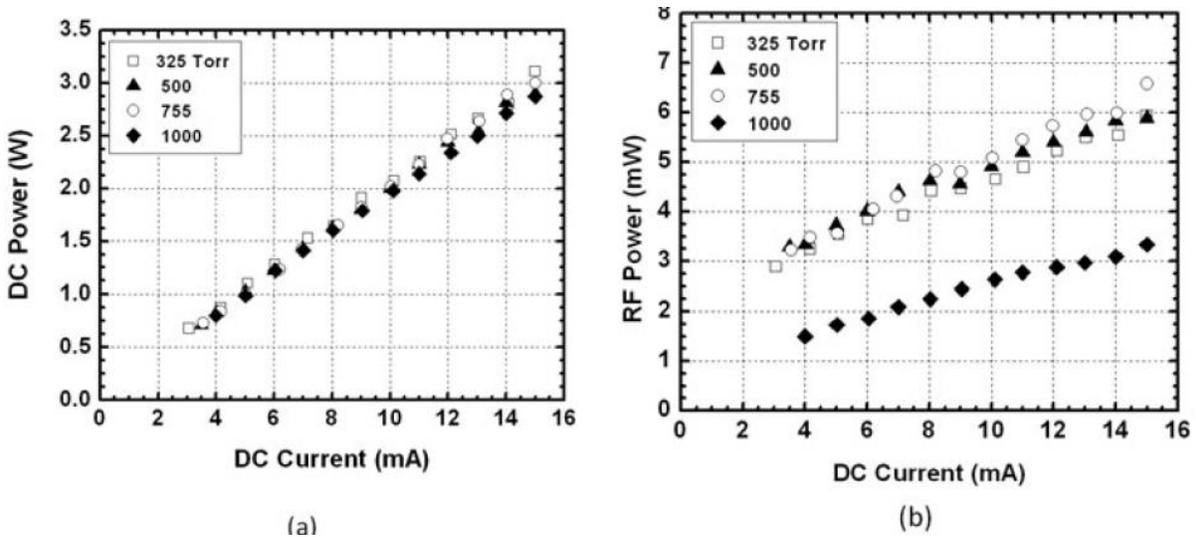

Fig. 2. Comparison of power deposited into the MHCD. (a) The DC power used to sustain the Argon plasmas as a function of the DC Current. (b) The RF power.

## 3. Results & Discussion

The DC micro-plasmas formed in this work have nearly constant voltage DC I – V curves, very similar to what one expects for normal glow discharges [8]. Examples of these can be seen in Figure 3. We note that Helium microplasmas are particularly susceptible to even very small amounts of impurities and believe that this could be the reason for the small variations in the Helium operating voltages at 250 Torr. Helium and Argon plasmas operate at approximately the same total voltages. Such constant voltage I – V curves indicate that the glow is operating in the so-called normal glow regime where the cathode sheath voltage is remaining nearly constant with increasing current and power because the cathode sheath area grows with increasing current.

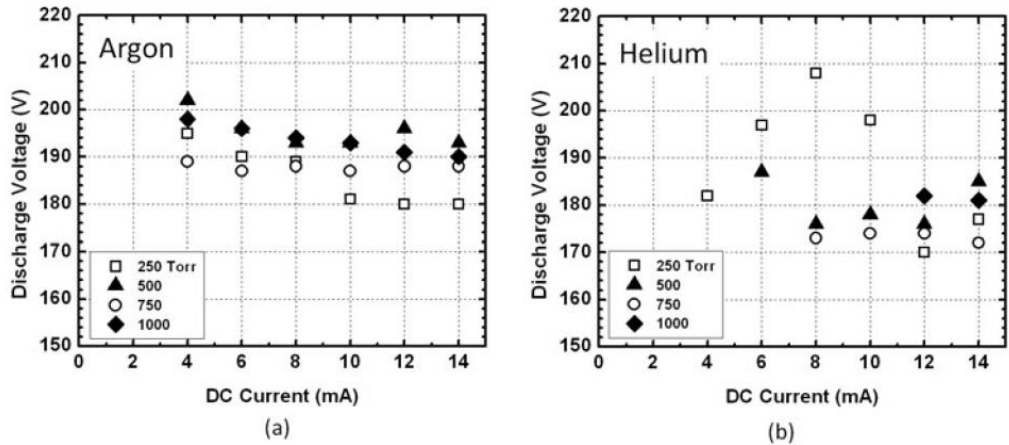

Fig. 3. Discharge I–V characteristics. (a) Argon. (b) Helium.

The positive-column like discharge in the through-hole has a very large collision frequency for electrons due to the large atom density ($v_m \gg \omega$) and causes the complex RF conductivity to be essentially real. (The inductive component becomes negligible.) Consequently, we model the net current flow in this region as a resistance, $R_p$. This is the plasma model shown in Figure 1a. In addition to $R_p$, there is the impedance of the plasma source itself, Cs. Cs is set by the area of the Ni electrodes on either side of the alumina insulator and the thickness of that insulator. The RF I–V probes allow one to measure the MHCD source's RF impedance and (if it were necessary) to deconvolve the imaginary and real parts. Consequently, after accounting for the effects of the tuning circuit, it is possible to obtain the micro-plasma's resistance at RF frequencies directly.

To analyze these measurements we start with the extremely well known formula for the conductivity of uniform plasma in conjunction with the equally well known relationship between conductivity and measured resistance. The RF measured resistance is then related to the plasma electron density by:

$$R_p = \frac{l}{\sigma A} = \frac{m \nu_m l}{n_e e^2 A_h}$$

where Ah is taken to be the area of the through-hole ($\pi r^2$) in the micro-plasma source for reasons to be discussed later, r is 90 μm, l is the length of that hole (500 μm), vm is the electron collision frequency for momentum transfer, m is the electron mass and ne is the average electron density in this (assumed uniform) plasma. We note that the measured resistance of the plasma is assumed to be dominated by impedance of the plasma inside the hole and not by the spread of the glow over the surface of the cathode. The plasma in the hole obviously has a much smaller and well defined cross-sectional area. The validity and impact of this assumption will be addressed later. Equation (1) has only a single unknown, and as a consequence, we can obtain an estimate of the electron density in the micro-plasma source directly from it (assuming that $v_m$ is known). The electron collision frequency is estimated as: $v_m = N_g K_{el}$ where $N_g$ is the gas density ($N_g = P/kT_g$) and the elastic collision rate constant $K_{el} = 5 \times 10^{-14}$ m³/s for argon [9,10] and $5.9 \times 10^{-14}$ m³/s for helium [10]. (These values correspond to an electron temperature $T_e \approx 1$ eV [1,11].) Estimating vm also requires knowledge of the gas temperature since the gas density (at a measured gas pressure) depends on the local value of $kT_g$. In subsequent figures we will plot results assuming that the gas temperature is 300 K and explain the difference which will occur if the gas temperature is larger. There is reason to believe that the gas temperature could be larger [1], however, the proximity of the alumina surface in the hole should help to moderate the gas temperature.

Since the RF I –V probes measure the impedance before the tuning circuit, it is helpful to note the procedure for obtaining the plasma impedance from the measured input impedance. The procedure is as follows: before turning the plasma on, RF is applied to the circuit and the tuning circuit is adjusted to provide maximum input impedance. This input impedance is typically 20 to 24 kΩ and occurs when the phase angle between current and voltage is minimized. It will be denoted by $Z_{NPin}$ (for input impedance with no plasma). Then the DC power supply is turned on and a plasma forms in the hole. The RF input impedance measured with plasma is always smaller than ZNP in and is denoted as $Z_{Pin}$. This $Z_{Pin}$ is the parallel combination of the RF plasma impedance that we want, $Z_P$, with the impedance of the circuit alone, $Z_{NPin}$. So the RF plasma impedance is found from:

$$Z_P = \frac{Z_{in}^{NP} \times Z_{in}^{P}}{Z_{in}^{NP} - Z_{in}^{P}}.$$

The phase angle of ZNP in is always set close to zero (since that is the point of maximum impedance) and in addition, the phase angle of ZP in is usually reasonably small too. In addition, the conductivity is expected to be real at this frequency, since the electron collision frequency is much larger than the probe RF frequency. As a result, it is reasonable to expect that the phase angle is not of particular importance in the present structure. In addition, we find that the phase angle measurements are not particularly stable in time. The measured phase angles just before striking the plasma and just after turning the plasma off ought to be the same, but are not the same even though the circuit has remained nominally constant. This indicates that the accuracy of the phase angle measurements are affected by the plasma through factors such as heat dissipation in the source structure (thermal expansion), possibly the erosion of the Ni film or other mechanisms. As a consequence, we have ignored the impedance phase angle here and simply utilized the magnitude of the input impedances to calculate the upper bound plasma resistance $R_p$ as the $|Z_P|$ from the $|Z_{NPin}|$ and $|Z_{Pin}|$. We do not believe this to be a major factor affecting the results but note that it simply gives us electron densities estimates which act as a lower bound.

In order to understand where the measurement of the plasma density is being made, we have plotted simulation results in Figure 4 of the electric potential contours in the micro-plasma source. Figure 4a is for a bulk plasma conductivity of 0.01 S/m and Figure 4b is for 10 S/m. The 2D axially symmetric simulation was performed using COMSOL's "AC/DC: Quasi-Statics, Electric" module and performed over a much larger volume than what is shown in Figure 4 in order to ensure that boundary conditions did not affect the results. This module allows one to solve Poisson's equation in a 2D axially symmetric configuration given dielectric permittivities and conductivities in a defined geometry. Figure 4 only shows the potential contours in the region of the MHCD source through hole since that is of primary concern. The micro-plasma structure consists of two Ni electrodes (solid black), the alumina insulator (between the Ni electrodes) and the plasma region highlighted in grey. There are two sections to the plasma region: the sheath over the upper Ni cathode (with a specified conductivity of $10^{-6}$ S/m) and the bulk plasma (0.01 or 10 S/m). The simulation did not calculate these plasma conductivities, instead that parameter was specified in order to get a first order estimate of the effect on the potential contours of an increasingly conductive plasma region. Toward that end, the 10 S/m value is larger than the conductivities found for the plasmas tested and corresponds to an electron density of approximately $5 \times 10^{14}$ cm$^{-3}$ at atmospheric pressure (assuming 300 K gas temperature). The 0.01 S/m value corresponds to a plasma density that is much lower than is stable in the hole (n ∼ $5 \times 10^{11}$ cm$^{-3}$). It is useful in that it shows essentially the same electric potential structure as is found when the conductivity is zero.

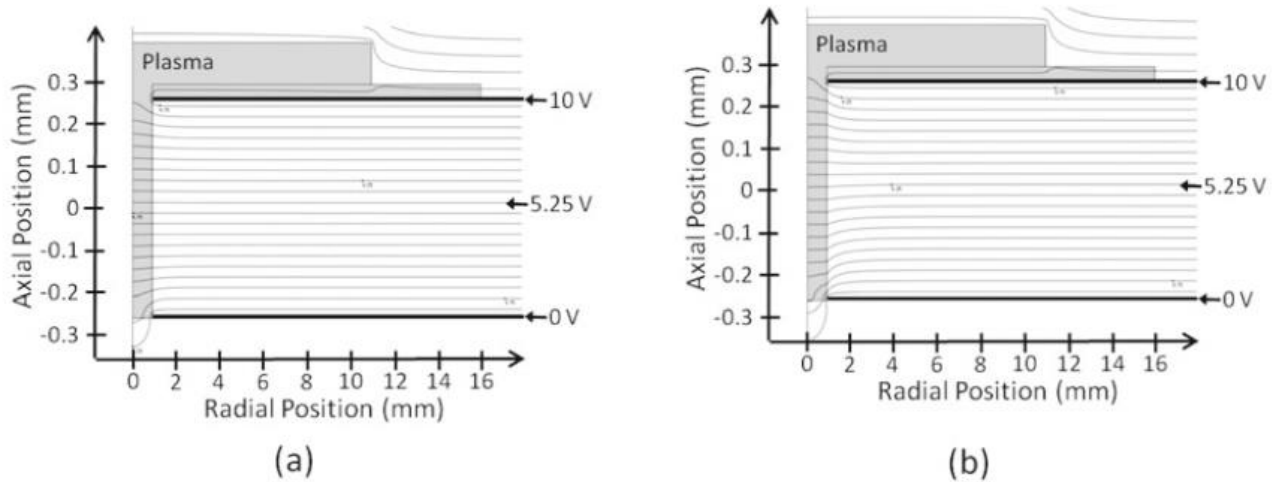

*Fig. 4. Equi-potential curves in the MHCD when the "plasma" conductivity is set to (a) 0.01 S/m (leaky insulator) and (b) 10 S/m (weak conductor).*

There are two significant features to note in this model. The first is that the potential contours are expelled from the plasma region above the upper Ni electrode. This would be the portion of the plasma that spreads over the cathode. As a result, little to no RF currents flow there. This clearly indicates that the measurements made using the RF do not measure the plasma density above the cathode. There is little to no field there for probing the plasma. On the other hand, there is a much stronger potential gradient in the hole and those potential contours remain nearly constant whether the plasma is strong (10 S/m) or extremely weak (0.01 S/m). This is evident from the fact that Figures 4a and 4b look almost identical. It highlights the most important outcome from the simulation. Namely: because the field in the hole is independent of the plasma density, the RF measurement will give information about the conductivity in the hole (and hence the electron density in the hole); not elsewhere. This is why we can use the cross sectional area and length of the hole in our calculations. While there is a slight difference in the potential contours between the two modeled conductivities (seen mostly at the lower Ni electrode region) this small difference should not cause significant errors in the calculations.

Figure 5 shows the magnitude of the plasma impedance, $Z_p$, measured by the RF I – V probes as a function of the DC discharge current and pressure. The plasma impedance ranges from ~30 kΩ to ~5 kΩ which results from an input impedance, $Z_{Pin}$, which is always significantly less than the no-plasma impedance of 20–25 kΩ. $Z_{Pin}$ also decreases monotonically with increasing DC current from roughly 12 kΩ to ~5 kΩ. This is as expected since the plasma density in the hole increases monotonically with current and thereby reduces $R_p$. On the other hand, there is no clear trend in the plasma impedance measurements with pressure. The measurements seem to group together indicating that the DC current level sets the plasma impedance nearly independently of the pressure.

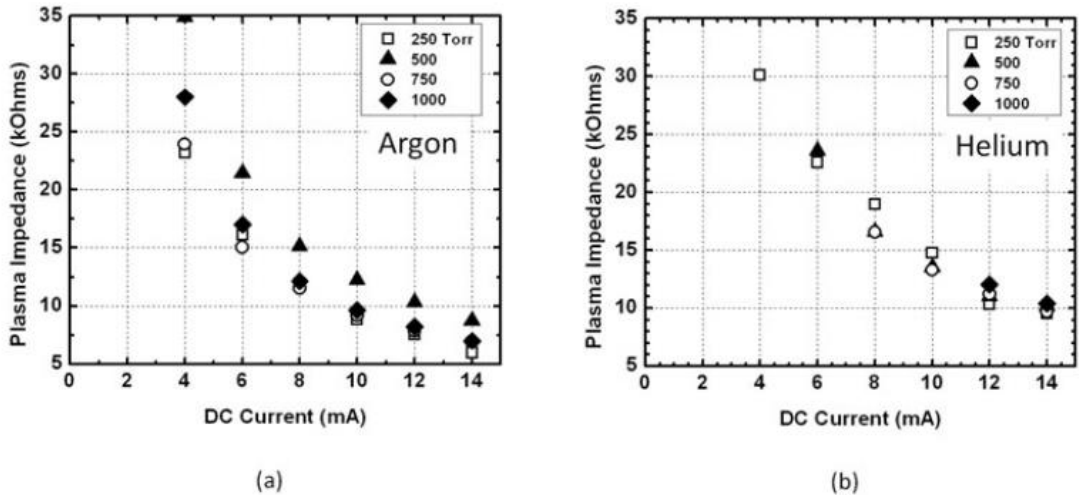

*Fig. 5. The measured plasma impedance in (a) Argon and (b) Helium MHCDs.*

Figure 6 is a plot of the electron densities estimated from the measured discharge resistance in argon as a function of discharge current. The electron density is found to rise with the DC current (and DC power since the voltage is almost constant). It has been assumed for these figures that the gas temperature is 300 K and constant. If the gas temperature is larger by some multiplicative factor, then the electron density will be reduced by the same factor since the collision frequency scales with the gas temperature through the gas density. We expect that the gas temperature could be larger than 300 K and dependent upon the discharge power as well as pressure, but we have no means of measuring it at present. In addition, the proximity of the alumina walls in the hole should act to moderate the gas temperature in the measurement region. Consequently we have assumed 300 K for the figures rather than speculate. Under this assumption, the estimated electron density in Argon rises from approximately $10^{13}$ cm$^{-3}$ to $1.7 \times 10^{14}$ cm$^{-3}$ as current and pressure rise. Helium electron densities are similar in magnitude, rising from $10^{13}$ cm$^{-3}$ to $1.3 \times 10^{14}$ cm$^{-3}$ in the same range of pressures and currents. These values appear to be somewhat lower than a fluid model and Stark broadening measurements on a similar MHCD [12]. The Stark broadening results were only obtained for helium plasmas, but indicated an electron density in 750 Torr plasmas ranging from $6 \times 10^{14}$ cm$^{-3}$ to $10^{15}$ cm$^{-3}$. The model results indicated an electron density peak of approximately $4 \times 10^{13}$ cm$^{-3}$ at 100 Torr and 5 mA current level. These are both significantly larger than the estimates from the RF measurements presented here.

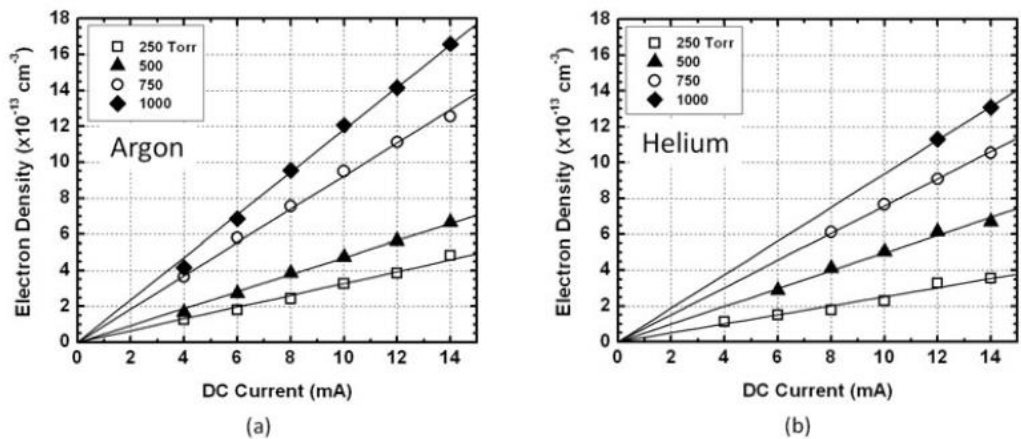

*Fig. 6. The estimated electron densities in (a) Argon and (b) Helium MHCDs as a function of the DC current level.*

There are a couple of potential explanations for the discrepancy. First, the fluid model indicated a significant sheath width in the plasma near the cathode [12]. The electron density in the sheath is much lower than that in the plasma bulk and as a result, the cross section of the plasma able to contribute to current flow is much smaller than the cross sectional area of the through hole and the plasma density in the center must be larger than estimated. The sheath width in the plasma was estimated by the model to be approximately 60 µm at 100 Torr [12]. Our MHCD hole has a radius of only 90 µm so this must be an upper bound for the sheath width at our pressures, but would be enough to cut the plasma cross sectional area by a factor of 9. Thus the electron density in the plasma would have to be approximately a factor of 9 times larger and much closer to the Stark measurements. We estimate that the actual sheath width along the insulating walls of the hole is more likely to be ~20 to 30 µm thick, which would indicate an increase by approximately a factor of 1.5 to 2. A second potential explanation is that the RF measurements estimate the average density along the length of the hole while the Stark broadening measurements see the peak density. It is highly likely that there is some significant variation in the density due to the structure of the MHCD.

The lines drawn in Figure 6 are linear fits showing that at each pressure the electron density rises nearly linearly under the assumption that the gas temperature is constant in the bulk of the glow at 300 K. If the gas temperature is rising, then the increase of the electron density would slow because the gas density would decrease, reducing the collision frequency. Thus, the linearity of the rise in the electron density is likely to be an overestimate of the actual electron density dependence. As the discharge current rises, the electron density should rise sub-linearly due to an increasing gas temperature and decreasing gas density in the through-hole. Note as well that the 250 Torr glow has a significantly smaller density than the 1000 Torr glow. This fits expectations since the collision frequency is larger at 1000 Torr making the electron mobility smaller and thus requiring larger numbers of electrons to carry the same DC current and current density. In addition, the diffusive losses of ions should be reduced at larger pressures, allowing a similar ionization rate (DC power) to result in a larger electron and ion density.

## 4. Conclusion

RF measurements of the impedance of DC MHCDs demonstrate a monotonically increasing plasma density with both the DC current level and pressure for helium and argon. Argon discharges exhibited a slightly larger plasma density than similar helium glows. Assuming the gas temperature in the hole remains nearly constant, the electron density rises linearly with discharge current. The RF impedance method is both inexpensive and less complex than other techniques for estimating this critical discharge parameter. The authors wish to acknowledge use of the UTD clean room in fabricating these devices.